\documentstyle[12pt,amssymb,epsf,apalike]{article}
\def\a{\alpha}
\def\b{\beta}

\def\s{\sigma}

\def\G{\Gamma}

\def\vp0{\varphi_{0}}

\begin{document}


\title{Surface-dependent Coagulation Enzymes: Flow Kinetics of
Factor Xa Generation on Live Cell Membranes}

\author{Maria P. McGee$^{*}$ and Tom
Chou$^{\dagger}$\\[72pt]
$^{*}$Department of Medicine \\
Wake-Forest University School of Medicine \\
Medical Center Blvd., Winston-Salem, NC 27157 \\[20pt]
$\dagger$Department of Biomathematics, UCLA School of Medicine \\
Los Angeles, CA  90095-1766}

\maketitle


\vspace{6cm}
\baselineskip=13pt






Running title: Flow kinetics of coagulation factor Xa

\newpage

\begin{center}
SUMMARY
\end{center}

The initial  surface reactions  of the extrinsic
coagulation pathway on live cell membranes  
were examined under flow
conditions.  Generation
of fXa (activated  coagulation factor X) was
measured on spherical monolayers of epithelial
cells with a total surface area of
41-47 cm$^{2}$ expressing TF(tissue factor)
at $>25$ fmol/cm$^{2}$. Concentrations of
reactants and product  were monitored  as a
function of time with radiolabeled  proteins and
a chromogenic substrate at resolutions of 2-8 s. 
At physiological concentrations of fVIIa and fX,
the reaction rate was $3.05\,\pm\,0.75$ fmol
fXa/s/cm$^{2}$, independent of flux, and 10 times
slower than that expected for collision-limited
reactions.  Rates were also independent of
surface fVIIa concentrations within the range 
0.6-25 fmol/cm$^{2}$.  The transit time of fX
activated on the reaction chamber was prolonged
relative to transit times of nonreacting tracers
or preformed fXa.  Membrane reactions were
modeled using a set of nonlinear kinetic
equations and a lagged normal density curve to
track the expected surface concentration of reactants
for various hypothetical reaction
mechanisms.  The experimental results were
theoretically predicted only when the models used
a slow intermediate reaction step, consistent with
surface diffusion.  These
results provide evidence that the transfer of substrate 
within the membrane is rate-limiting in the kinetic
mechanisms leading to initiation of blood coagulation by the
tissue factor pathway.  

\vspace{2mm}

\begin{center}
INTRODUCTION
\end{center}

Blood coagulation  reactions  mediate  fibrin
deposition in hemostasis and 
many pathological processes.  Blood clots are
directly implicated  in the lethal complications
of cardiovascular disease  and contribute
significantly to the pathogenesis of
infectious, autoimmune, and neoplastic diseases
(1-6).

The blood coagulation process is initiated by
an assembly of complexes comprised of an essential
cofactor, TF (tissue factor) and a protease
component, fVIIa.  The functional complex,
TF/fVIIa,  cleaves the natural substrates, fVII,
fIX (factor IX), and fX  at specific sites,
generating, fVIIa, fIXa, and fXa, respectively
(4-6).  Factors VII, IX, and
X circulate in the blood and extravascular
fluids (7-10), while TF
is expressed on the membranes of many
extravascular tissues \cite{DRA89}.  The
anatomic distribution of cells expressing TF is
consistent with its role as the initiator of
hemostatic reactions.  Cell surfaces in contact
with blood do not appear to express functional
TF constitutively. However, inflammatory
stimuli induce expression of
functional TF on endothelial cell membranes and
blood monocytes (12-14).

Factors VII, IX, and X are vitamin K-dependent
proteins, and their functional interaction with
negatively charged procoagulant membranes has a
calcium-dependent, electrostatic component
(15-19). The interaction sites are located in 
highly homologous $\gamma-$carboxyglutamic acid (Gla)
rich regions near the N-terminus of all vitamin K-dependent
coagulation proteins (4, 19). 
The specific binding and functional kinetics of
interaction between coagulation proteins and
biological membranes have been studied
extensively under equilibrium steady-state
conditions (20-27).  While equilibrium binding
parameters vary significantly among vitamin
K-dependent proteins, adsorption parameters are
similar, suggesting nonspecific initial contact
(28, 29). Anionic phospholipid
membranes modify the apparent kinetic parameters
of coagulation reactions relative to kinetics in
solution. The membrane effect is manifested by a
large decrease in the apparent $K_{m}$ of
substrates to values far below their respective
plasma concentrations. The mechanisms by which
this effect manifests itself during TF-mediated
coagulation remain speculative. Achieving useful
time resolutions has been one of the main
obstacles to developing experimental systems to 
study the presteady-state transients of coagulation
factor adsorption and activation on cell membranes.
Blood clotting {\it in vivo} and {\it in vitro}
can be completed faster than the sampling
intervals of traditional batch systems used to measure 
membrane reactants.

Measurements of fVIIa binding and fXa generation
on intact cell membranes under steady-state
conditions indicate that TF/fVIIa functional
activity is fully expressed before the binding
interaction between fVIIa and TF reaches equilibrium
(24, 25).  Furthermore, under steady-state
conditions, the overall rate of coagulation substrate
activation on membranes pre-equilibrated with enzyme
was close to the theoretical collisional limit
\cite{MCG92a}. These findings suggest
intermediate noncovalent steps on the membrane
linking  the initial adsorption step  to the assembly
and catalysis of substrate in the A-E-S
(activator-enzyme-substrate).


Coagulation zymogens and active proteases are subject
to local microcirculation controls (3, 7-10, 14) since
they are found in extravascular lymphatic, synovial,
and alveolar fluids. The importance of flow control
in coagulation reactions has been demonstrated {\it
in vivo}. Tracer studies with radiolabeled fibrinogen
and vasoactive agents indicate a direct correlation
between changes in vascular permeability and fibrin
deposition \cite{DVO85}. Several studies using
lipid-coated capillaries  also indicate  that flow
rates influence the activity of coagulation proteases
(30, 31). In the present study, we use
high-resolution tracer-dilution analyses (32-35)
along with numerical modeling to identify the surface
and flow-dependent kinetics of fX activation via the
TF pathway.  We show that the generation of fXa from
plasma fX proceeds via an intermediate step within
the membrane.  For reactions initiated  with fVIIa
and fX, the rate of this step and of the overall
reaction is limited  by the transfer of fX from the
adsorption sites to the catalytic sites on the cell's
surface.  


\begin{center}
EXPERIMENTAL PROCEDURES
\end{center}

{\it Cell Culture and Reaction Chambers}--Cells and cell
cultures have been described and characterized in
detail elsewhere (29, 34).  Briefly, Vero
cells (American Tissue Type Collection) were grown to
confluency  on microcarrier beads (Cytodex 2,
Pharmacia) of 150 $\mu$m average diameter.  Reaction
chambers were assembled with a 2-3 ml suspension of
cell-covered microspheres loosely packed in a
thermoregulated column fitted with flow adapters. A
schematic of the reaction flow chamber is shown in
Figure \ref{SPHERE}.  

Cell viability and metabolic integrity, as demonstrated by amino
acid uptake, were maintained for periods exceeding the kinetic
measurements described here \cite{PER86}. The cells expressed
surface-TF constitutively and interacted functionally with human
fVIIa and fX with apparent steady-state kinetic  parameters
similar to other procoagulant cells, as reported previously (29,
35).  The TF activity of the monolayer suspension in the reaction
chamber was equivalent to $6149\pm 847$ fmol/ml of recombinant TF
reconstituted into phospholipid vesicles. The chambers were
perfused with medium (M-199, Gibco), buffered with HEPES, and
supplemented with 0.1 mM of ovalbumin and 3 mM CaCl$_{2}$.  The
overall geometrical parameters and TF content in the reaction
chamber are summarized in Table I.

{\it Tracer-dilution techniques}--Concentrations of reactants
and products in the bulk, flowing, aqueous phase and on the
cell surface during fXa-generating reactions were measured
using double and triple-tracer techniques. These techniques
were adapted from previously described and validated methods
in perfused organs (32-34, 36, 37).  Coagulation initiation
reactions were carried out at physiological concentrations,
under flowing conditions.  Reactions were initiated under flow
by injecting 48-500 ng of fVIIa and 3000-9000 ng of fX
in a 1.23 ml reaction chamber \footnote{These amounts 
of reactant were
rapidly (1-2 s) injected in bolus of 100 $\mu\ell$}.
The injected reactants are quickly dispersed in the reaction
chamber and attain maximal initial concentrations of
approximately 0.8-8 nM fVIIa and 45-135 nM fX.  
concentrations but diminish over time.  Control tracer 
$^{14}$C-labeled ovalbumin (833 ng) was also included 
in the bolus injection yielding an
initial  maximal concentration of approximately 14 nM.  To
measure membrane concentrations,  $^{3}$H-labeled fVIIa was
used an adsorbing tracer.  Reactants and control 
tracer were in a perfusing medium containing 0.1 mM unlabelled 
ovalbumin and 3 mM CaC$\ell_{2}$.

After introducing reactants into the chamber
via the inflow tubing, the effluent  was collected
in 72 samples of $46\pm 2.8\,\mu$l each at time
resolutions ranging from 2-10 s per sample,
depending on the perfusion rate.  At flow rates of 13
$\mu$l/s, the typical flow velocity in the reaction
chamber was estimated at $\lesssim 0.2$ mm/s.  This value
is within the ranges measured in rabbit
microvasculature {\it in vivo} \cite{SCH75}.  

Samples were collected in microtiter wells
preloaded with TRIS buffer solution, pH 8.3,
containing 0.2 M EDTA and 0.8 M NaCl. A final 5 ml
sample was collected  to complete the recovery 
of tracer and to determine perfusion rates. Standard
curves for $^{14}$C and $^{3}$H tracers were
constructed from serial dilutions of the solutions
used in the bolus injection. Concentrations of
control and test radioactive tracers in standard
dilutions and effluent samples were measured by
scintillation counting. The concentration of
product, fXa, was measured by amidolytic assays (13, 14)
with chromogenic substrate
(Methoxycarbonyl-D-cyclohexylglycil-arginine-p-n
itroanilide-acetate) before scintillation counting
of radioactive tracers.  

{\it Functional tests for TF Activity in
Cells}--Functional activity  of TF was determined
from fXa generation rates in purified systems using
amidolytic  assay and recombinant proteins as
standard.  Recombinant TF used to construct standard
curves was relipidated into 30:70 PS/PC
(phosphatydylserine/phosphatydylcholine) vesicles as
before (17, 31).  The molar ratio of
TF/lipid in the standard TF-PS/PC preparations was
1/4100.  The TF activity measured in intact
monolayers was $68.5\pm 19\%$ of that measured in
monolayers lysed  by freezing/thawing.

{\it Radioactive Tracers}--The control
tracer used to measure concentrations in bulk
aqueous phase was $^{14}$C-labeled ovalbumin
(Sigma, St. Louis, MO) with a  specific activity
of 33 $\mu$Ci/mg.  The test tracer for
adsorption measurements was fVIIa radiolabeled
with tritium using the technique of Van Lenten
and Ashwell \cite{VAN71}, with modifications
\cite{SAN85}.  Labeled preparations had specific
activities of 1.8$\times$10$^{8}$ cpm/mg of
fVIIa  and a functional activity comparable to
unlabeled factor in clotting tests and
activating mixtures with purified components
\cite{MCG99}.  

{\it Reaction scheme and mathematical
model}--The surface reactions leading to fXa
generation  were analyzed according to the
following scheme:

\begin{equation}
\begin{array}{cc}
\mbox{fVIIa} + \mbox{TF} \begin{array}{c} 
k_{+E} \\[-5pt] 
\rightleftharpoons
\\[-5pt] k_{-E}\end{array} \mbox{E} \\[13pt]
\mbox{E}+\mbox{fX} \begin{array}{c} k_{+a}
\\[-5pt] 
\rightleftharpoons
\\[-5pt] k_{-a}\end{array} \mbox{E}\!\cdot\!\mbox{fX}
\stackrel{k_{+}}{\rightarrow}\mbox{fXa}+\mbox{E},
\end{array}
\label{SCHEME}
\end{equation}

\vspace{3mm}

\noindent where $\mbox{E}\equiv
\mbox{fVIIa}\!\cdot\!\mbox{TF}$ is the fVIIa and TF
complex (``enzyme'') that forms and
dissociates with rate constants $k_{+E}, k_{-E}$,
respectively. The substrate-enzyme complex denoted
by E$\cdot$fX associates and dissociates with a
second-order rate-constant, $k_{+a}$, and a
first-order rate-constant, $k_{-a}$, respectively. 
The effective rate of product (fXa) formation from
the complex and its irreversible release are 
denoted by the first-order rate-constant, $k_{+}$.  
Denoting the surface concentrations of each species
by $(\G_{1},\G_{2},\G_{3},\G_{4},\G_{5},
\G_{6})\equiv (\mbox{fVIIa, fX, TF, E,
E}\!\cdot\!\mbox{fX}, \mbox{fXa})$, the
full kinetic equations consistent with
(\ref{SCHEME}) are

\begin{equation}
\begin{array}{l}
\dot{\Gamma}_{1} =
-k_{+E}\G_{1}\G_{3}+k_{-E}\G_{4}-
\b_{1}\G_{1}+\a_{1}C_{1}(t)\\[13pt]
\dot{\Gamma}_{2} = -k_{+a}\G_{2}\G_{4}+
k_{-a}\G_{5}-\beta_{2}\G_{2}+\a_{2}C_{2}(t)\\[13pt]
\dot{\Gamma}_{3} = -k_{+E}\G_{1}\G_{3}+k_{-E}\G_{4}
\\[13pt]
\dot{\Gamma}_{4} = k_{+E}\G_{1}\G_{3}-k_{-E}\G_{4}-
k_{+a}\G_{2}\G_{4}+(k_{-a}+k_{+})\G_{5} \\[13pt]
\dot{\Gamma}_{5} =k_{+a}\G_{2}\G_{4}-(k_{-a}+k_{+})\G_{5}\\[13pt] 
\dot{\Gamma}_{6} =k_{+}\G_{5}+\alpha_{6}C_{6}-\beta_{6}\G_{6},
\end{array}
\label{GAMMADOT}
\end{equation}

\vspace{2mm}

\noindent where $\dot{\Gamma}_{i}(t)\equiv
d\Gamma_{i}(t)/dt$, with $i=(1,2,3,4,5,6)$.  The 
constants $k_{\pm E}$ and $k_{\pm a}$ correspond to
effective rates of TF and fX binding interactions with TF
and E respectively. These coefficients include the  time
delays of all intermediary processes  on the membranes
before the interactions. The time distributions of these
unspecified processes are accounted for in our  numerical
predictions of the time course for the overall fX
activating process.  The possibility of inhibition or
fX/fXa-destroying sinks is precluded from our data since,
within experimental error, all the absorbed fX is
recovered as fXa. 



In the above nonlinear differential equations, $\b_{i}$ and
$\a_{i}$ are desorption and adsorption rates, respectively. 
Since under our experimental conditions the total area
fraction of adsorbed species is negligible, species adsorption
from bulk is simply proportional to the bulk concentration,
$C_{i}(t)$, at the surface of each microsphere. As the fluid
passes through the ensemble of microcarriers, certain flow
lines are faster or slower than the mean flow velocity,
resulting in a distribution of reactant velocities. A ``lagged
normal density curve'' (LNDC) has been successfully used to
approximate the dispersion resulting from the combined effects
of random velocity distribution and molecular diffusion in the
human circulatory system (36, 37).  We find good agreement
between a fitted lagged density curve and the sequentially
measured concentrations in the outflow of the reaction chamber
(Fig.  \ref{ControlfX}).  Therefore, to simplify the modeling
process, we assume that the dispersion and diffusion of all
species are equal and use the LNDC to approximate the source,
$C_{i}(t)$, surrounding each microcarrier. The parameters used
in the lagged density curve will reflect chamber packing
characteristics, bulk diffusion constants, and the imposed
volume flow rate, $J_{V}$.  Additional details, analysis,  and
simplifications of Eqs.  (\ref{GAMMADOT}) are provided in the
Appendix.

{\it Calculation of reactant concentrations in membranes}--The
proportion of fX and fVIIa adsorbed from the flowing phase into
the membrane was determined from the difference between the
normalized concentrations  of control, $^{14}$C-labeled
ovalbumin, and $^{3}$H-labeled fVIIa.  Concentrations of factor
VIIa adsorbed at time $t$ were estimated using

\begin{equation}
\Gamma_{1}(t) \approx \left( 
\left[^{14}\mbox{C}(t)\right]-\left[^{3}\mbox{H}(t)
\right]\right)Q_{T}S_{T}^{-1},
\end{equation}

\noindent where $\left[^{14}\mbox{C}\right]$ and
$\left[^{3}\mbox{H}\right]$ are the fraction of
the total nonadsorbed control and adsorbed test
tracer, respectively, collected in the effluent.
$Q_{T}$(in fmol) is the total amount of fVIIa
added, and $S_{T}$ is the total membrane surface
area ($\approx$ 41-47 cm$^{2}$).  Previous studies
measuring adsorption of various coagulation
factors, including fVIIa and fX, indicated that
adsorption rates are proportional to their
aqueous-phase concentration and not
significantly different among vitamin K-dependent 
proteins (28, 29).  Based on these data, the
normalized concentration of $^{3}$H-labeled
fVIIa was used  to trace
both fVIIa and fX adsorption.  For
substrate fX, the membrane concentration
$\Gamma_{2}(t)$ was determined by

\begin{equation}
\Gamma_{2}(t) \approx
\left(\left[^{14}\mbox{C}(t)\right]-
\left[^{3}\mbox{H}(t)\right]-\left[\mbox{fXa}(t)
\right]\right)Q_{T}S_{T}^{-1}, 
\end{equation}

\noindent where $Q_{T}$ is the total amount
(fmol) of fX added, and
$\left[\mbox{fXa}(t)\right]$ is the fraction of that
total released into the
aqueous phase as fXa.

{\it Miscellaneous}--Tissue factor antigen
expressed by the cells was determined in cell
lysates using a commercial ELISA kit with
recombinant soluble TF as standard (American
Diagnostics).  Protein determinations were
performed in the same cell lysates with a
commercial reagent (Biorad Laboratories), using bovine serum albumin as standard. 
Coagulation fVIIa was human recombinant, kindly
donated by Dr.  Ulla Hedner (Novo Nordisk,
Denmark).  Recombinant tissue factor (used as
standard in ELISA and functional TF determination
in cells) was purchased from American Diagnostics.
Human fX and fXa were purchased from Enzyme
Laboratories. Data reduction, plotting, and
statistical analyses were performed using
StatView software (Brain Power, Inc.).  Numerical
solution of the nonlinear kinetic equations
(\ref{GAMMADOT}) was performed using adaptive
Runge-Kutta implemented through Matlab.

\vspace{2mm}

\begin{center}
RESULTS
\end{center}

{\it Rate of fXa generation at different fVIIa
concentrations}--The generation of fXa from fX on live
procoagulant cell membranes was examined in reaction
chambers filled with spherical cell monolayers.  The
geometrical  and flow characteristics of these
reaction chambers are summarized in Figure
\ref{SPHERE} and Table I, respectively.  The
distribution of concentrations of reactants  in the
flowing bulk aqueous phase was followed using control
tracer $^{14}$C ovalbumin.  Reactions were initiated
with fX and fVIIa, and the product, fXa, was measured
by amidolytic assay. Reactions
were followed until 70-90\% of the nonreacting control
tracer was recovered in the effluent. The amounts of
$^{14}$C and $^{3}$H tracer collected and the amount
of $^{3}$H adsorbed to the cell are shown as
functions of time in Figure \ref{TRACER}A.

The time evolution of reactants inside the
reaction chamber can be fairly well approximated by
the LNDC, as shown in
Figure \ref{ControlfX}. This agreement indicates that
dispersion of reactants in the chamber due to
the random flow distribution and diffusion is
qualitatively similar to that encountered in 
human circulation (36-38, 43, 46-47). Figure
\ref{ControlfX} also illustrates the
time/concentration distribution of fXa generated
in the 1.23 ml reaction chamber in a typical reaction initiated
with 500 ng of fVIIa and 9000 ng of fX.

Under these conditions, aqueous-phase
concentrations ranged from 0.3-10$\,\pm\,2$ nM and
from 4-137$\,\pm\,$29 nM for fVIIa and fX,
respectively. A time trace measuring the total
amount of fXa collected and fX
adsorbed on cell membranes is shown in Figure
\ref{TRACER}B. Factor Xa profiles were weakly
sigmoidal with a linear middle segment. Table II 
shows that the
average production rate, calculated from the
linear segment, did not change when average
concentrations of fVIIa in aqueous phase were
decreased by ten-fold, from 5 nM to 0.5 nM 
(membrane concentrations ranged from 0.7-25.0
fmol/cm$^{2}$). No fXa was generated in the absence of 
fVIIa and
reaction rates did not differ significantly
when fVII was substituted for fVIIa.  The
observation that maximal constant catalytic
activity is reached at very low concentrations of
TF$\cdot$fVIIa complexes allows for simplifying 
substitutions in the model equations (Eqs.
\ref{GAMMADOT}) for $\G_{1}$ and $\G_{4}$
(Appendix).
 
{\it Comparison between the reaction rate
and the theoretical collisional rate}--The
independence of average reaction rates and
enzyme concentrations suggests that substrate
transfer to the catalytic sites is rate
limiting.  Two possibilities were investigated:
{\it (i)} the rate-limiting step may occur
during the adsorption of reactants from bulk to
the cell membranes, and {\it (ii)} the 
rate-limiting step occurs after the adsorption step. 
To differentiate these two
possibilities, the rate of fXa generation was
compared to the theoretical collisional rate between 
reactants and microspheres, given the
aqueous phase concentrations of fX used and the
flow rates, $J_{V}$, imposed. Since average
adsorption rates were shown to approach or to
exceed the collisional limit \cite{MCG99},
activation rates below this limit support the
second possibility.

Theoretical steady-state collisional rates were
calculated from the aqueous-phase concentrations
of fX and the radius of the spherical
microcarriers  using  Smoluchowski's 
relationship for steady-state diffusion (42, 57)

\begin{equation} 
k_{\em coll} \approx D_{1,2} C_{1,2}(t) R^{-1},
\label{COLLISION}
\end{equation}

\noindent where $k_{\em coll}$ is the collision rate
between reactant molecules and a unit area of membrane
(collisions/cm$^{2}$/s), $D_{1,2}$ is the diffusion constant
for fVIIa, fX in water ($\sim 5\times
10^{-7}\,\mbox{cm}^{2}\mbox{/s}$), $R\simeq 7.5\times
10^{-3}\,\mbox{cm}$ is the microcarrier radius, and $C_{1,2}$
is the fVIIa, fX concentration (molecules/cm$^{3}$).  

Figure \ref{ControlfX} contrasts the number of fXa
molecules released by the monolayer and the
aqueous phase concentration, $C_{2}(t)$, of fX as a
function of time.  Since the collisional rate follows
Eq. \ref{COLLISION}, collision-limited rates are
expected to be directly proportional to $C_{2}(t)$.
However, the rate of fX activation on the monolayer
was not correlated with fX-membrane collisions. The
rate of fXa production (molecules/cm$^{2}$/s)
reached maximal values after the peak in $C_{2}(t)$
and collisional rates.  Furthermore, high fXa rates
were sustained during the rapid decrease in
collisions between fX and the membrane, following
the concentration peak.  Averaged over 13
experiments, the activation rate was $3.05\pm 0.72$
fmol/cm$^{2}$/s, corresponding  to $1.8\,\pm\,
0.43\times$10$^{9}$(molecules/cm$^{2}$/s), below
the theoretical maximum of $2.4\pm 0.57
\times$10$^{10}$(collisions/cm$^{2}$/s).  

Apparent second-order rate coefficients,
calculated from initial  reaction rates and
aqueous-phase concentrations  of fVIIa and fX,
did not have a constant value but increased
continuously during the observation time. 
Using aqueous-phase fX and membrane-phase  fVIIa
to calculate  apparent second order rate
coefficients also resulted in increasing
coefficient values, consistent with the
observation that the average reaction rates are
essentially independent of fVIIa concentration
(Table II).  These results indicate that the
initial adsorption from bulk to membrane is not
rate-limiting in the overall fXa generation
reaction.  The results also imply that fX is
activated  via a membrane-bound intermediate
rather than directly from the bulk aqueous phase.

{\it Mean transit times of fX through the
reaction chamber}--The existence of a slow
membrane step was further investigated by
comparing average transit times of the fXa
generated in the chamber to the transit times
of preformed fXa.  The presence of a
rate-limiting step between membrane
adsorption and catalytic cleavage is expected to
delay the transit of the substrate that is
adsorbed and catalyzed as compared to bulk
aqueous-phase reactants.  A mean transit time,
$T_{D}$, was determined from the concentration of
fXa and control tracer in 72 consecutive samples
of the effluent according to the expression

\begin{equation}
T_{D} = {\sum_{n=1}^{72}
\left[^{*}\mbox{C}(t_{n})\right]\times t_{n}
\over \sum_{n=1}^{72}
\left[^{*}\mbox{C}(t_{n})\right]}, 
\label{TIME}
\end{equation}

\noindent where
$\left[^{*}\mbox{C}(t_{n})\right]$ is the
fraction (of the total amount added or
theoretical maximum) of either control tracer or
fXa collected in aliquot $n$ at time $t_n\simeq
n\times (2-8)$ sec, depending on the 
particular experiment.  Results shown in Table III
indicate that the mean transit time of fX, 
$T_{D}(\mbox{fX})$, activated in the reaction
chamber is increased relative to the $T_{D}$ of
aqueous-phase control tracer. In contrast,
$T_{D}(\mbox{fXa})$ for fXa formed before
being introduced in the reaction chamber is
indistinguishable from that of the control
tracer.  Furthermore, the increase in
$T_{D}(\mbox{fX})$ is inversely correlated with
flow rate. These results are consistent with a
slow membrane step following the fast,
flow-dependent adsorption step.

{\it Kinetic modeling of surface reactions}--The
hypothesis that the reaction pathway proceeds with
fast equilibration of enzyme activity followed by a
rate-limiting step involving reactant
surface diffusion was also tested by comparing
experimental measurements to the solutions of the
kinetic equations (\ref{GAMMADOT}).  Equations
(\ref{GAMMADOT}) were solved numerically using
initial estimates for intrinsic rate constants
based upon results of previous steady-state kinetic studies
(15, 20, 21, 25-27). Heuristic arguments for initial guesses
for all the rate parameters are provided in the
Appendix. A continuous function for aqueous-phase
concentrations $C_{i}(t)$ is derived from a 
least-squares fit to a LNDC (Appendix), shown in Fig.
\ref{ControlfX}.  The remaining parameters in the
model were then adjusted until the best
visual fit of $\Gamma_{6}(t)$ to  fXa
collected was achieved. Since $\Gamma_{2}(t)$ was
indirectly measured and subject to larger
experimental errors, we only varied rate parameters
to get an order-of-magnitude agreement between the
measured fX (Fig. \ref{ControlfX}) and
$\Gamma_{2}(t)$ (Fig. \ref{FIG4}B), using measured
fXa and $\Gamma_{6}(t)$ to more precisely fit the
parameters.  The solutions and the associated best-fit 
parameters are shown in Fig.  \ref{FIG4}. We found
that the magnitudes of $\Gamma_{i}(t)$ match 
the measurements only when the amount of TF
assumed in the simulations was 0.32 fmol/cm$^{2}$,
much smaller than the actual amount expressed on the
cell membranes. This finding is consistent with our
hypothesis that enzyme complexes form domains, 
further developed in the Discussion Section. 

The model also shows that within reasonable ranges, 
the shape and magnitude of the product curve, 
$\Gamma_{6}(t)$, are sensitive to $k_{+a}, k_{+}$ and 
$\a_{i}, \b_{i}$, but less sensitive to the 
other parameters.  If the association step, $k_{+a}$,
were fast, the theoretical model would predict a
premature overproduction of fXa, as shown in Figure
\ref{FIG4B}A. The sensitivity to a slow intermediate
membrane step associated with $k_{+a}$ is shown in
Fig.  \ref{FIG4B}C, while the corresponding predicted
values for effective enzyme, TF, and fX on the
membranes are shown in Fig. \ref{FIG4B}D. 
Note also that for parameters differing from those
used in Fig. \ref{FIG4}, the magnitudes of 
$\Gamma_{2}(t)$ change dramatically and are no longer 
close to the measured fX (Fig. \ref{ControlfX}).
 
For the parameters used to fit the 
measurements in Fig. 2, the numerical solution for E
($\Gamma_{4}$) plateaus to a value 
$\sim\Gamma_{4}^{*}$ after $t\sim 40$ s and remains
nearly constant for the duration of
the 300 s interval under consideration.  This
quasi-steady-state exists even for the cases
where $k_{+a}$ is too large (Figure
\ref{FIG4B}B) or too small (Figure
\ref{FIG4B}D). We show in the Appendix that
this quasi-steady-state behavior allows us to
define an approximate {\it effective} rate
constant (s$^{-1}$)

\begin{equation}
k_{\em eff}\equiv {k_{+}k_{+a}\G_{4}^{*} \over 
k_{+}+k_{-a}}
\label{KEFF}
\end{equation}

\noindent that approximately describes the rate
of fXa production on the cell membranes via 

\begin{equation}
\dot{\Gamma}_{6} \approx k_{\em
eff}\Gamma_{2}(t)+\mbox{adsorption/desorption
terms}.
\end{equation}

For large $k_{+}$ (the fast chemical step), $k_{+a}$
becomes the limiting rate, since $k_{\em
eff}\approx k_{+a}$. The experimental data are
consistent with model predictions  both
qualitatively and quantitatively, when $k_{+a}$  is
in the range expected for lateral diffusion of
proteins on membranes. An estimate for a 
mechanistically relevant diffusion length can be 
derived from 

\begin{equation}
\ell_{D} \sim \left({D_{2}^{\em surf}
\over k_{\em eff}}\right)^{1/2},
\end{equation}

\noindent where $D_{2}^{\em surf}$ is the
surface diffusion constant of fX in the cell
membranes.  Using a typical value $D_{2}^{\em
surf} \sim 10^{-10}\,\mbox{cm}^{2}$/s
\cite{CHE79}, and $k_{\em eff}\sim
0.02\,\mbox{s}^{-1}$ obtained from the
mathematical model, we find that $\ell_{D} \sim
0.7\,\mu$m. However, $\ell_{D}$ can be shorter, if
obstructions in the membrane hinder surface
diffusion and reduce $D_{2}^{\em surf}$.  

{\it Factor Xa generation rate as a
function of surface density of reactants}--The
observations described in the previous sections
indicate that reaction rates are not directly
related to the aqueous-phase concentration of
substrate.  To further investigate  the
rate-limiting step we analyzed  reaction rates
as a function  of flow rates and membrane
concentrations of substrate. 


The instantaneous fraction of membrane fX
converted to fXa was not directly proportional
to the fX concentration on the membrane. 
Instead, at all flow rates tested, it increased
linearly with time. Interestingly, the increase
in the proportion of adsorbed fX encountering
catalytic sites per unit time was essentially
independent  of flow. This observation suggests that
adsorbed fX is not immediately available to the
catalytic sites and is consistent with a 
rate-limiting surface diffusion process.  That the
fraction of membrane fX available for catalysis
increases with time suggests diffusive
transfer of fX from initial adsorption sites to
the catalytic sites.  The value of the slope of
the function $\G_{6}(t)/\G_{2}(t)$ was 0.001 -
0.0005 $s^{-1}$ and largely independent of flow
rate (Table IV).
 
Table IV also lists the product yield and
average reaction rates measured at different
flow rates.  The yield was strongly correlated
with flow rate (correlation coefficient 0.86),
while the average reaction rate was independent
of flow rate. Again, these results are as
expected for a kinetic mechanism that includes a
flow-dependent adsorption step followed by a
rate-limiting intermediate step on the membrane.

\vspace{2mm}
\begin{center}
DISCUSSION
\end{center}

In this paper, we have analyzed the surface reaction
kinetics of blood coagulation initiated by
interaction among coagulation factors on
biological membranes.  Reactions were initiated
on live epithelial cells expressing TF with 
physiological concentrations of fVII/fVIIa and fX. 
The aqueous and membrane concentration of reactants
as well as product formation were measured at
time resolutions relevant to plasma clotting. 
Whereas adsorption of vitamin K-dependent
proteins on procoagulant membrane surfaces has
been shown to be fast and correlated to
aqueous-phase flux \cite{MCG99}, the rate of fXa
generation was independent of both enzyme
density on the membrane and flow rate, $J_{V}$. 
Moreover, using tracer dilution analyses we
found that the transit time of the fX
participating in the reaction was prolonged
relative to transit times of nonreacting control
tracers. Rate coefficients calculated
from reaction rates and either the aqueous or
membrane concentration of reactants
changed with time. Flow velocities influenced the total amount
of reactants adsorbed to the 
membrane and the total yield, but not the 
intrinsic rate of product formation. Taken together,
the experimental results provide evidence for a
kinetic mechanism limited by a slow transfer of
substrate between initial membrane adsorption
sites and reaction sites.  The hypothesis of a
slow membrane step was further tested by
numerically solving a set of nonlinear kinetic
equations describing the evolution of all
membrane reaction species. The experimental
results were reproduced when the rate-limiting
step followed the adsorption of substrate to the
membrane  and preceded the chemical catalysis. 
Testing the alternative hypothesis of either
slow adsorption or slow catalytic steps resulted
in product yield and profiles markedly different
from those observed experimentally.  

Maximal surface catalytic activity was observed
when the membrane concentration of fVIIa was at
least two orders of magnitude  lower than the
surface concentration of TF, estimated  either by
immunoassay or functional tests.  The
experimentally measured fXa levels best
matched solutions of the kinetic equations
(Eqs. \ref{GAMMADOT}) when the intrinsic rate
constants shown in Fig. \ref{FIG4} were used
along with a maximal enzyme concentration of
$\Gamma_{4}\simeq 0.32$ fmol/cm$^{2}$. This
observation suggests that only a fraction of the
available membrane E = fVIIa$\cdot$TF is involved
in catalysis as would be expected if catalytic
sites were in large molar excess over the
substrate.  However, the substrate concentration
reached, $>100$ fmol/cm$^{2}$, is much higher
than TF or enzyme concentrations. 

These results can be explained using a model in which the
enzyme concentration, $\Gamma_{4}$, is in relative local
excess, and only a fraction of the surface enzyme effectively
participates in catalysis. If enzyme complexes fVIIa$\cdot$TF
form domains, the fast catalytic cleavage reaction would be
expected to occur only near the perimeter of these enzyme
domains, where the substrate initially encounters the enzyme
after diffusing a typical distance, $\ell_{D}$, following
adsorption. The interiors of these domains are rarely accessed
by the rapidly converted substrate and therefore do not
participate in the overall reaction.  This process is shown
schematically in  Figure \ref{cluster}.  

Surface segregation of molecules is a common
phenomenon. It has been shown that even
small molecules can form domains in lipid
monolayers (44, 48, 49) and bilayer
vesicles \cite{BAG00}. A large body of
experimental research also provides 
evidence that protein and lipid domains
exist on live cell membranes
(50, 51).  For example, adhesion
molecules localize at sites of cell-cell
contact, and receptors are often found
concentrated at the tips of filopodia and
lamelopodia of moving cells. Neurotransmitter
receptors in postsynaptic terminals have also
been shown to form dynamic aggregates
(53, 54). Direct visualization
reveals localization of certain proteins in
areas of high or low curvature in artificial
vesicles \cite{DISCHER}. We have evidence of
the existence of domain formation
on the membrane of the epithelial cell line used
in these studies.  Using gold immunochemistry on
cells fixed after short a exposure to fVIIa, the
enzyme was localized primarily on the ruffled
border of the cell membrane. Furthermore,
analysis of nearest-neighbor distances
indicated a nonrandom distribution of the enzyme
\cite{MCG99}.  Although the mechanisms of domain
formation are unknown, possibilities may involve
electrostatic or dipole-induced phase
transitions \cite{AND}, and membrane 
elasticity-induced protein-protein 
attractions \cite{KIM}.

The flow characteristics and time resolution 
achieved with this experimental system is
relevant to reactions on biological membranes
after exposure of TF to flowing plasma
coagulation proteins. The experimental
approaches and mathematical model used here 
to identify the early kinetic mechanisms of fXa
generation will also be useful for studying 
novel pharmacokinetic mechanisms occuring
on biomembranes.

\vspace{3mm}
\begin{center}
APPENDIX
\end{center}

In this section, we give details of the
mathematical model (Eqs. \ref{GAMMADOT}) and the
associated approximations used in its analysis. 
Since no measurable amount of fXa is generated
by fVIIa and fX in bulk solution, we have
assumed that chemical reactions can only occur
when molecules are adsorbed on the surface of
each cell-covered sphere. 

The kinetic equations (Eqs. \ref{GAMMADOT}) are
solved numerically using finite difference
approximations.  All surface densities,
$\Gamma_{i}$, are in units of fmol/cm$^{2}$,
while all bulk concentrations, $C_{i}$, are
measured in units of pmol/cm$^{3}$. 
With this convention, the rates take on the
following units:
$[\beta]=[k_{-a}]=[k_{-E}]=[k_{+}]=\mbox{s}^{-1}
$, $[k_{+E}]=[k_{+a}]=\mbox{cm}^{2}/(\mbox{fmol
s})$, and $[\alpha]=10^{-3}\mbox{cm/s}$.

We assume that the adsorption rates of species
from the bulk onto the microsphere surfaces are
proportional to the local bulk concentration. 
For the sake of completeness, and to motivate
more quantitative modeling, we
write the governing equations for surface
adsorption of reactants. In the bulk phase,
the concentration of species {\it i} follows the
convection-diffusion equation:

\begin{equation}
\partial_{t}C + {\bf V}\cdot\nabla
C = D_{i}\nabla^{2}C 
\quad \quad r \geq R,
\label{CONVDIFF}
\end{equation}

\noindent where $C = C_{1}, C_{2}, C_{6}$, and $D_{i}$ are the
bulk solution concentrations (number per volume) and
associated diffusion constants of fVIIa, fX, and fXa,
respectively.  Although a closure relation is required 
to specify the detailed velocity field, ${\bf V}$, around
each sphere, we will assume that the identical microspheres
each feel an equivalent, averaged, effective flow velocity,
${\bf V}$.  The boundary conditions within continuum
theory at the sphere surface are found by balancing the
diffusive flux with the desorption and adsorption rates on the
surface of each microsphere (at $r=R$),

\begin{equation}
D_{i}\partial_{r} C_{i}(r=R, t) =
\a_{i} \Phi C_{i}(R, t)
-\b_{i}\Gamma_{i}(t),
\label{BCAB0}
\end{equation}

\noindent where the area fraction available for
adsorption $\Phi\simeq 1$ at low coverage.  

The above equations constitute the exact
continuum equations for the species in solution.
The complete set of equations consists of the
convection-diffusion equation (Eq. 
\ref{CONVDIFF}), the boundary conditions (Eqs. 
\ref{BCAB0}), and the surface reaction equations
(Eqs. \ref{GAMMADOT}). 

Significant simplifications and decoupling of
some of Eqs. \ref{GAMMADOT} can be realized by assuming 
that the bulk concentrations at the
spheres' surfaces $C_{i}(R,t)$ can be approximated
by the LNDC \cite{BAS66}. As the fluid passes
through the ensemble of cell-covered
microspheres, certain flow lines are faster or
slower than the mean flow velocity.  The
reactants in the aqueous phase are randomly
advected through the microsphere chamber at a
distribution of velocities. The LNDC is a
convolution of random advection velocities with
molecular diffusion and has been used to
approximate advection-diffusion in blood flow
\cite{BAS66}.  Although the source concentration,
$C_{i}(t)$, may depend on the position of the
microsphere within the reaction chamber, we 
assume that the dispersion and diffusion of all
species are equal and use the lagged density
curves to approximate the source, $C_{i}(t)$,
surrounding each microsphere. The parameters
used in the LNDC depend upon average
microsphere packing, the bulk diffusion
constants, and the imposed constant volume flow
rate, $J_{V}$.

The concentration of the $i^{th}$ species in a
random flow environment is assumed to obey

\begin{equation}
\tau_{i} {d C_{i}(R,t) \over d t} + C_{i}(R,t) = {1000 m_{i}
\over J_{V} \left(2\pi
\sigma_{i}^{2}\right)^{1/2}}
\exp\left(-{1\over 2}\left[(t-T_{i})/\sigma_{i}\right]^{2}\right)
\label{BOLUS}
\end{equation}

\noindent where $m_{i}$ is the total number of femtomoles of
species $i$ added via the bolus injection, and $J_{V}$ is the
constant flow rate measured in $\mu l/s$. The intrinsic delay
time, $T_{i}$, is inversely related to the mean $\vert {\bf
V}\vert$, while $\sigma_{i}$ and $\tau_{i}$ describe the width
and effects of molecular dispersion, respectively.  The amount
of spreading embodied in $\sigma_{i}$ is proportional to the
bulk diffusion constant, $D_{i}$. The solution to the initial
value problem (Eq. \ref{BOLUS}) is

\begin{equation}
C_{i}(t)= {1000 m_{i} \over 2\tau_{i}J_{V}} e^{\s_{i}^{2}/2\tau_{i}^{2}}
e^{-(t-T_{i})/\tau_{i}}\left[\mbox{Erf}\left({\sigma_{i} 
\over \sqrt{2}\tau_{i}}
+{T_{i}\over 
\sqrt{2}\sigma_{i}}\right)-
\mbox{Erf}\left({\sigma_{i} \over \sqrt{2}\tau_{i}}-{(t-T_{i})\over 
\sqrt{2}\sigma_{i}}\right)\right].
\label{C_i}
\end{equation}

The concentration, $C_{i}$, above is given in
nM units. These solutions determine the sources,
$\alpha_{i}C_{i}(R,t)$, for the surface kinetic
equations (\ref{GAMMADOT}).  We take the entire
reaction chamber and the inlet and outlet tubes
to constitute a single, effective flow system. 
The zero used in Eq. \ref{C_i}
corresponds to the time when nonbinding species are
first detected (for the experiment in Fig. 
\ref{ControlfX}, approximately 36 s after adding
reactants).  Upon fitting (by adjusting
$\sigma_{i},\tau_{i}, T_{i}$ until a local minimum
in the least-squares is found)  the parameters
in Eq. \ref{C_i} to the concentration, we find
(for this experiment at $J_{V}\simeq 13.4\,\mu$l/s)
that $\sigma_{i}\approx 21$ s, $\tau\approx
91$ s, and $T\approx 19$ s for $i=1,2$. The
fitted LNDC is shown in Fig.  \ref{ControlfX}.  

The initial rate parameters used to solve Eqs. 
\ref{GAMMADOT} were estimated as follows.  The
absorption rates of fVIIa and fX, from previous
studies, were found to be similar and close to
the collisional limit \cite{MCG99}.  Here, we
first assumed that these adsorption rates were
diffusion-limited. Such
high absorption rates were needed to
obtain the right magnitudes of fXa formation,
regardless of the other rate parameters. Enough
reactant must simply reach the membranes within
the time limit imposed by the flow rate, $J_{V}$.
The maximum rate of particles reaching and
absorbing into the membrane of an isolated
microsphere's surface, in the absence of flow, is
given by 

\begin{equation}
J_{i} \leq 4\pi
R D_{i} C_{i}(r=\infty,t).
\end{equation}

\noindent This upper limit assumes that every
molecule coming into contact with the sphere is
absorbed.  With a diffusion constant of
$D_{i}\sim 7.5\times 10^{-7}$ cm$^{2}$/s, the
absorption rate under zero flow conditions is
approximately $\alpha_{i} \lesssim 1000 D_{i}/R
\approx 0.1$ cm/s (the factor 1000 converts
pmol/cm$^{3}$ to fmol/cm$^{3}$).  Now consider
the effects of advection due to the imposed
volume of flow, $J_{V}$.  Purcell \cite{PURCELL}
gives an expression for the flux to a spherical
surface under flow:

\begin{equation}
J_{i}({\bf V}) \leq 4\pi R D_{i} C_{i}(t)
\left({R V\over D_{i}}\right)^{1/3}.
\end{equation}

\noindent Although at first glance the $V^{1/3}$
dependence is weak, quantitative changes in
$\alpha$ due to flow can dramatically influence
the yield of the surface reactions. For the
density of microspheres and flow rates used, we
estimated the typical velocity in the reaction
chamber to be $V \approx J_{V}/A_{\em eff}\approx
0.02$ cm/s, where $J_{V}\simeq 13.4\,\mu$l/s, 
and the reaction chamber effective
cross-sectional area $A_{\em eff} < \pi
(0.7\,\mbox{cm})^2$ due to partial 
obstruction.  Therefore, we used
$\alpha_{1,2}\sim 0.6$ cm/s as an initial guess
for the absorption rates in the reaction scheme
(Eqs. \ref{GAMMADOT}).

For the association rates $\alpha_{i}$, we used
the surface diffusion of fVIIa and fX to set
upper limits. For a surface diffusion constant of
$D_{i}^{\em surf} \sim 10^{-10}$ cm/s, we found that
$k_{+a}\sim k_{+E} < 0.1 \, \mbox{cm}^{2}$/(fmol
s).  Moreover, from previous aqueous-phase equilibrium binding
studies, $k_{-E}/k_{+E} \sim 10^{-10}$ M (24-26). 
To estimate the corresponding ratio for
two-dimensional reactions, we made a 
{\it qualitative} estimate by assuming the 
energetics of E formation are not 
significantly different from those in bulk. 
Thus, $10^{-10}$ M  corresponds to 
a typical particle-particle distance of 
2.5 $\mu$m. Translating this to a 
surface density, we estimated very roughly
that for the surface enzyme formation reaction, 
$k_{-E}/k_{+E} \sim  
10^{-2}$ fmol/cm$^{2}$. Finally, from
previous steady-state measurements on cell
membranes at saturated concentrations of fX,
$k_{+}\gtrsim 14$/s \cite{PLO67}.  
Within these limits, we explored
the parameter space to obtain a reasonable fit
to the data.  The data and the fit of
$\Gamma_{6}(t)$ calculated from (\ref{GAMMADOT})
are shown in Fig.  \ref{FIG4}A. The
concentration of product collected, $Q_{6}(t)$, is
calculated from 

\begin{equation}
Q_{6}(t) \approx \beta_{6}\Gamma_{6}(t) S_{T}/J_{V},
\label{Q6}
\end{equation}

\noindent where $S_{T}\simeq 47\, \mbox{cm}^{2}$ is
the total membrane area in the reaction chamber.
The membrane concentration $\Gamma_{2}$ is shown
in Fig. \ref{FIG4}B.

The results were consistent with the data in Table II
and showed
that after a short initial transient, the
concentration of membrane enzyme ($\G_{4}$)
reaches a plateau.  This behavior permitted simplification
of Eqs. \ref{GAMMADOT} and 
approximate analytic solutions for the surface
concentrations $\G_{2,6}(t)$. For large
$k_{+}+k_{-a}$, the
concentration $\G_{5}(t)$ (E$\cdot$fX) is always
small.  From a typical simulation Figs. 
\ref{FIG4}, \ref{FIG4B} we observed a short
transient in $\G_{3}$ (TF). At times beyond this
transient, $\dot{\G}_{3}\approx 0$, and the
enzyme concentration, $\G_{4}$, reaches a nearly
steady value:

\begin{equation}
\G_{4}^{*} 
\approx {k_{+E}\over k_{-E}}\G_{1}\G_{3}.
\end{equation}

From the equation for $\dot{\G}_{1}$, we see
that over long times, fVIIa approximately follows
the adsorption and desorption processes,

\begin{equation}
\dot{\G}_{1} \approx \alpha_{1}C_{1}(t)-\beta_{1}\G_{1}.
\label{GAMMA1}
\end{equation}

\noindent It is
evident from Fig. \ref{FIG4}-\ref{FIG4B} that $\G_{4}$ (E)
also reaches a quasi-steady state shortly after
TF. Therefore, setting $\dot{\G}_{4}\approx 0$,

\begin{equation}
\dot{\G}_{5} \approx {k_{+a}\G_{4}^{*} \over k_{+}+k_{-a}}\G_{2}
\equiv k_{\em eff}\G_{2}. 
\end{equation}

\noindent The remaining time-dependent surface quantities 
at these quasi-steady-state times obey 

\begin{equation}
\left(\begin{array}{c} \dot{\G}_{2}(t) \\ \dot{\G}_{6}(t)
\end{array}\right) = \left(\begin{array}{cc}
-k_{\em eff} & 0 \\
k_{\em eff} & -\beta_{6} \end{array}\right)\left(
\begin{array}{c} \G_{2}(t) \\ \G_{6}(t)\end{array}\right) 
+ \left(\begin{array}{c}
\alpha_{2}C_{2}(t) \\ \alpha_{6}C_{6}(t)\end{array}\right),
\label{2X2}
\end{equation}

\noindent where $k_{\em eff}$ given by Eq.
\ref{KEFF} is the effective rate of conversion
from fX to fXa during quasi-steady-state times
when the enzyme concentration is $\G_{4} \approx
\G_{4}^{*}$. Assuming that $\a_{6}C_{6}$ is
negligible, Eqs. \ref{2X2} admit analytic
solutions and, considering the 
approximate nature of our model, a further
simplification can be made: Using Eq. \ref{Q6}, 
the second equation in (\ref{2X2})
becomes

\begin{equation}
\dot{Q}_{6} = 
k_{\em eff}\beta_{6}\left[
{S_{T}\over J_{V}}\Gamma_{2}-{1\over k_{\em eff}}
Q_{6}\right].
\end{equation}

Therefore, an independent measurement of $\Gamma_{2}$ and the
collected product $Q_{6}$ can be used to estimate the unknowns
$\beta_{6}$ and $k_{\em eff}$.  Although in our analyses we
have numerically solved the full kinetic equations (Eqs. 2), a
simplified set of equations (Eqs. \ref{GAMMA1} and \ref{2X2})
provide an analytic model to the reaction kinetics for times
beyond the initial short transient ($t\gtrsim 40$ s for the
run shown in Fig. \ref{FIG4}).

\vspace{3mm}

\noindent {\it Acknowledgments}--This work was
supported by NSF grants MCB-9601411,
DMS-9804370, and NIH-HL57936

\vspace{3mm}

\begin{center}
REFERENCES
\end{center}

\newpage

\begin{figure}[t]
\begin{center}
\leavevmode
\end{center}
\caption{\baselineskip=12pt
{\bf Schematic of the reaction chamber
and cell-covered microspheres.}  The reactive
surface in the reactor is the surface of viable
Vero cells grown to confluency on microcarriers
with the indicated dimensions.  The
microcarriers  ($\sim 5-20\times 10^{4}$) are
packed in a thermoregulated column fitted with
flow adapters and perfused at constant flow
rates of $5-25\,\mu$l/s. Reactants and control
tracer are added via the inflow (lower) and
collected via the outflow (upper) tubing in
72-140 consecutive samples of $46\pm 8.9\,\mu$l
each. For most experiments, reactants are added
as a rapid bolus and reactions followed for
150-300 s by collecting samples at a resolution
of 2-10 s per sample.}
\label{SPHERE}
\end{figure}

\begin{figure}[t]
\begin{center}
\leavevmode
\end{center}
\caption{\baselineskip=12pt
{\bf Adsorption of reactants and factor Xa generation
under flow.}  The reaction chamber was maintained at 37 C
and perfused at 13.4 $\mu\ell/s$ with HEPES buffered
medium, pH 7.2, containing 0.15 N NaCl, 3 mM
CaC$\ell_{2}$, and 0.1 mM nonlabelled ovabumin.  Maximal
initial concentrations of reactant were 8 nM $^{3}$H-fVIIa
and 130 nM fXa.  The TF density on the monolayer surface
was estimated at $>25$ fmol/cm$^{2}$ from both functional
and immunological assays.  (A) Total amounts of reactant
(either fVIIa or fX, $\Box$) adsorbed to the monolayer
were determined from the difference between the normalized
concentrations of control tracer, $^{14}\mbox{C}
(\bigcirc)$, and test tracer,
$^{3}\mbox{H}(\blacktriangle)$, collected in effluent
samples.  Tracer amounts were normalized as the fraction
of the total added to the reaction chamber. (B) The amount
of fX on the membrane $(\bigcirc)$ was determined from the
difference between fX adsorbed $(\Box)$ and fXa
$(\bullet)$ released.  The average rate of fXa generation
for this experiment calculated from the slope of the
middle linear segment of the progression curve (100-150
sec) was $2.1\pm 0.02$ fmol fXa/s/cm$^{2}$.  The mean from
13 similar experiments was $3.05\pm 0.72$ fmol/s/cm$^{2}$.}
\label{TRACER}
\end{figure}


\begin{figure}[t]
\begin{center}
\end{center}
\caption{\baselineskip=12pt {\bf The lagged normal density curve
and distributions of substrate and product
concentrations}. The open circles correspond to
concentrations of fX (in $nM$) in aqueous phase
determined from the concentration of
nonreacting, nonadsorbing control tracer.  The
zero of the time axis is chosen to correspond to initial
detection of $^{14}\mbox{C}$.  The qualitative
fit to the lagged normal density curve (LNDC)
yields the parameters $\sigma_{2}\approx 21$ s,
$\tau_{2}\approx 91$ s, and $T_{2}\approx
19$ s. The filled circles correspond to the fXa
concentration released into each aliquot in the
{\it reacting} system ($\times 10$ in the figure
to facilitate comparison with fX values).} 
\label{ControlfX}
\end{figure}


\begin{figure}[t]
\begin{center}
\end{center}
\caption{\baselineskip=12pt {\bf Kinetic modeling of surface
reactions}. The set of nonlinear equations (Eqs.
\ref{GAMMADOT}) was solved numerically and the
associated parameters adjusted to obtain the
best visual fit between computed and
experimentally measured fXa concentration
curves.  (A) Bulk fX concentration
($0.1C_{2}(t)$, black) and product fXa
($\G_{6}$, red). The approximate parameters achieving the
best fit are: $k_{+}=15, k_{+a}=0.06, k_{-a}=6,
k_{+E}=0.06, k_{-E}=0.0005, \alpha_{1,2}=0.8,
\alpha_{6}=0.1,\beta_{1}=\beta_{2}=0.001$, and
$\beta_{6}=0.12$. The amount of TF present in 
accessible enzyme complexes was assumed 
to be 0.32 fmol/cm$^{2}$.
(B) Corresponding surface concentrations
$\Gamma_{2}/100$, $\Gamma_{3},\Gamma_{4},\,
\mbox{and}\,\,\Gamma_{5}$ in fmol/cm$^{2}$.}
\label{FIG4}
\end{figure}

\begin{figure}[t]
\begin{center}
\end{center}
\caption{\baselineskip=12pt {\bf Model predictions for
alternative reaction mechanisms}. (A) The predicted fXa
generation as a function of time if the surface diffusion is
much faster ($k_{+a}=1.0$) than that assumed in the
simulations depicted in Fig. \ref{FIG4}, with all other
parameters identical to those used in Fig. \ref{FIG4}. (B) The
corresponding surface concentrations. Note that membrane fX,
$\Gamma_{2}$, is much smaller than that in Fig \ref{FIG4} and
estimated from measurements (not shown). (C) The predicted fXa
production if the surface diffusion were slower than optimal.
With $k_{+a} = 0.01$, factor Xa is generated in 
lower quantities and at later times.  (D) The predicted
$\Gamma_{2}$ however, is much greater than that 
observed.}
\label{FIG4B}
\end{figure}

\begin{figure}[t]
\begin{center}
\end{center}
\caption{\baselineskip=12pt {\bf Cartoon representing  
protein distribution on the cell
surface}.  If the formation of E$\cdot$fX occurs upon
nearly each encounter of E and fX on the membrane surface, 
only E molecules near the perimeters of the domains will
participate in catalysis of fX$\rightarrow$fXa.}
\label{cluster}
\end{figure}

\vspace{6mm}



\begin{center}
\begin{table}[htb] $\:$ \hspace{4.3cm} \mbox{TABLE
I}\\[4pt].\quad\quad\quad\quad\quad\quad
{\em Reaction\,\, Chamber\,\,Parameters}\\[1.5ex]
\begin{tabular}{ll} \hline\\[0.7ex]
Dimensions (length$\times$diameter)  & 0.8 cm$\times$1.4 cm  \\[1ex]
Volume & 1.23 cm$^{3}$ \\[1ex] 
Microcarriers/chamber & $6.7\pm 0.2\times 10^{4}$ \\[1ex]
Microcarrier radius & $7.5\times 10^{-3}$ \\[1ex]
Microsphere area & $7.0\times 10^{-4}\,\mbox{cm}^{3}$ \\[1ex]
Microcarrier/chamber volume fraction & $\sim 13\%$ \\[1ex]
$^{a}$Total reactive surface $S_{T}$ & $\sim 47\,\mbox{cm}^{2}$\\[1ex]
$^{b}$TF antigen & 0.63 fmol/$\mu$g protein \\[1ex]
TF activity & 2.3 fmol/$\mu$g protein \\[1ex]
\hline
\end{tabular}

\vspace{3mm}
\noindent $^{a}$Total surface estimated from
microcarrier counts in samples from 3
different chambers.

\noindent $^{b}$Tissue factor antigen was measured by ELISA
in detergent-lysed monolayers. Activity equivalent 
was estimated from TF activity in suspensions of intact
and lysed (by one freeze/thaw cycle) cell monolayers, 
relative to the activity of standard dilutions of human
recombinant TF reconstituted in phospholipid vesicles
(30:70, PS/PC). The TF activity in suspensions of intact
cell monolayers was $68.5 \pm 19\%$ of that in
suspensions of lysed monolayers.

\end{table}
\end{center}


\begin{center}
\begin{table}[htb]$\:$\hspace{5.2cm} \mbox{TABLE
II}\\[4pt].\quad\quad\quad\quad\quad\quad\quad
{\em fXa\,\,generation\,\,rate\,\,as\,\,
function\,\,of\,\,fVIIa}\\[1.5ex]
\begin{tabular}{lllr} \hline\\[0.7ex]
$^{a}$Enzyme density \quad \quad   &  $\:\quad\quad^{b}\mbox{Rate}$ &
$\:\quad\quad^{c}$Flux(fmol/s)& $\:$ \\[1ex]
(fVIIa fmol/cm$^{2}$) & (fmol fXa cm$^{-2}$s$^{-1}$)
\quad\quad & fVIIa & fX  \\[1ex]
5.0-25 & $2.9\pm 0.05$ & 180 & 3241 \\[1ex] 
2.2-9.0 & $3.7\pm 0.04$ & 52 & 779 \\[1ex]
1.9-3.4 & $3.9\pm 0.08$ & 23 & 1728 \\[1ex]
0.7-1.8 & $3.7\pm 0.06$ & 11 & 608 \\[1ex]
0 & 0 & - & 2000\\[1ex]
\hline 
\end{tabular}

\vspace{3mm}
\noindent $^{a}$ Membrane density and adsorption rate 
of fVIIa were derived from the flux and  
adsorption rate coefficients previously measured
\cite{MCG99}. Values are the initial and final
concentrations measured during the linear interval of
the progression curve. Tissue factor density in these
experiments was estimated at $>25$ fmol/cm$^{2}$ from 
cell protein and specific activity assays.

\noindent $^{b}$ Average rate was calculated from the
concentration of fXa measured by chromogenic assay in
samples collected during the linear segment of the
reaction progression curve.

\noindent $^{c}$ Average fluxes of fVIIa and fX in each
experiment were determined from the concentration of
$^{14}\mbox{C}$-control tracer in the effluent. The
total amounts of fVIIa and fX initially added to the 1.23 ml
reaction chamber were 48-500 ng and 
2800-9000 ng, respectively.

\end{table}
\end{center}


\begin{center}
\begin{table}[htb] \hspace{4.8cm} \mbox{TABLE III}
\\[4pt].\,\quad\quad\quad\quad\quad\quad
{\em Mean\,\,transit\,\,times\,\,
of\,\,controls\,\,and\,\,\mbox{fXa}}\\[1ex]
\begin{tabular}{c|rrrr} \hline \\[1pt]
Factors & \underline{Flow rate $\mu$l/s} & \multicolumn{3}{c} 
\mbox{\underline{Mean transit time $T_{D}$ (seconds)}} \\[1ex]
$\:$ & $\:$ & control\quad & fXa \quad\quad\quad & $\%$ change \\
\hline \\
$\:$ & $21.4 \pm 2.3\quad$ & $56\pm 5\quad$ & $79 \pm 5.0\quad$ & 
$41 \pm 10$\\[1ex]
$^{a}$X, VIIa & $11.7 \pm 1.0\quad$ & $107 \pm 12\quad$ & 
$144 \pm 23\quad$ & $32\pm 7.0$ \\[1ex]
$\:$  & $6.6 \pm 0.7\quad$ & $247 \pm 63\quad$ & $270\pm 68\quad$ & 
$12 \pm 3.0$ \\[1ex]
\hline \\
$\:$ & 21.0\quad\quad\quad & 54\quad\quad\quad & 54\quad\quad\quad 
& 0\quad\quad\quad \\[1ex]
$^{b}$Xa & 10.0\quad\quad\quad & 109\quad\quad\quad 
& 107\quad\quad\quad & -1\quad\quad\quad \\[1ex]
$\:$ & 8.0\quad\quad\quad & 106\quad\quad\quad 
& 167\quad\quad\quad & 1\quad\quad\quad \\[1ex]
\hline
\end{tabular}

\vspace{3mm}
\noindent $^{a}$ Mean transit times, $T_{D}$, of fXa and 
$^{14}\mbox{C}$-ovalbumin were calculated using Eq. 
\ref{TIME}. Factors VIIa and X were added with 
$^{14}\mbox{C}$-tracer, and fXa was generated in the
chamber. Values are from four experiments at each flow rate.

\noindent $^{b}$ Preformed fXa was added with 
$^{14}\mbox{C}$-tracer. Values are from one experiment
at each flow rate.

\end{table}
\end{center}


\begin{center}
\begin{table}[htb] $\:$\hspace{6.0cm}\mbox{TABLE
IV}\\[4pt].\,\, \quad\quad\quad\quad\quad
\quad\quad\quad\quad
{\em Effects\,\,of\,\,flow\,\,rate\,\,
on\,\,reaction}\\[1.5ex]
\begin{tabular}{rcrc} \hline \\[0.7ex] 
$J_{V}(\mu\mbox{l/s})$  &$^{a}d(\G_{6}(t)/\G_{2}(t))/dt\,(\times
10^{3}\,s^{-1})$ 
& $^{b}$Yield (\%) 
& Average rate(fmol cm$^{-2}$ s$^{-1}$)  \\[1ex]
$4.6 \quad $ & $0.5\pm 0.02$ & $17.6 \quad $ & 2.75 \\[1ex]  
$6.0 \quad $ & $1.0\pm 0.05$ & $17.5 \quad $ & 3.21 \\[1ex]
$6.4 \quad $ & $2.0 \pm 0.09$ & $17.1 \quad $ & 2.29\\[1ex]
$13.3 \quad $ & $1.0 \pm 0.04$ & $7.8 \quad $ & 2.29 \\[1ex]
$14.2 \quad $ & $1.0 \pm 0.05$ & $5.2 \quad $ & 2.34 \\[1ex]
$27.0 \quad $ & $1.0 \pm 0.15$ & $6.0 \quad $ & 2.85 \\[1ex]
$28.0 \quad $ & $1.0 \pm 0.04$ & $3.9 \quad $ & 2.91 \\[1ex]
\hline
\end{tabular}

\vspace{3mm}
\noindent $^{a}$ The fraction of membrane-bound fX
converted to fXa increased linearly with time giving
the slopes (roughly 0.001 $s^{-1}$) indicated. 
The density of fX on the membrane
was calculated from the difference between fX adsorbed
and fX released as fXa. The concentration of the fXa
released was measured directly in the effluent samples
by amidolytic assay.

\noindent $^{b}$ The yield of fXa is expressed as the
percentage of the total fX added to the reaction
chamber. Average reaction rates were determined from
the steepest linear segment of the progression curves
as in Fig. \ref{TRACER}B.  

\end{table}
\end{center}

\clearpage
\baselineskip=12pt

\begin{figure}[htb]
\begin{center}
\leavevmode
\epsfxsize=3.6in
\epsfbox{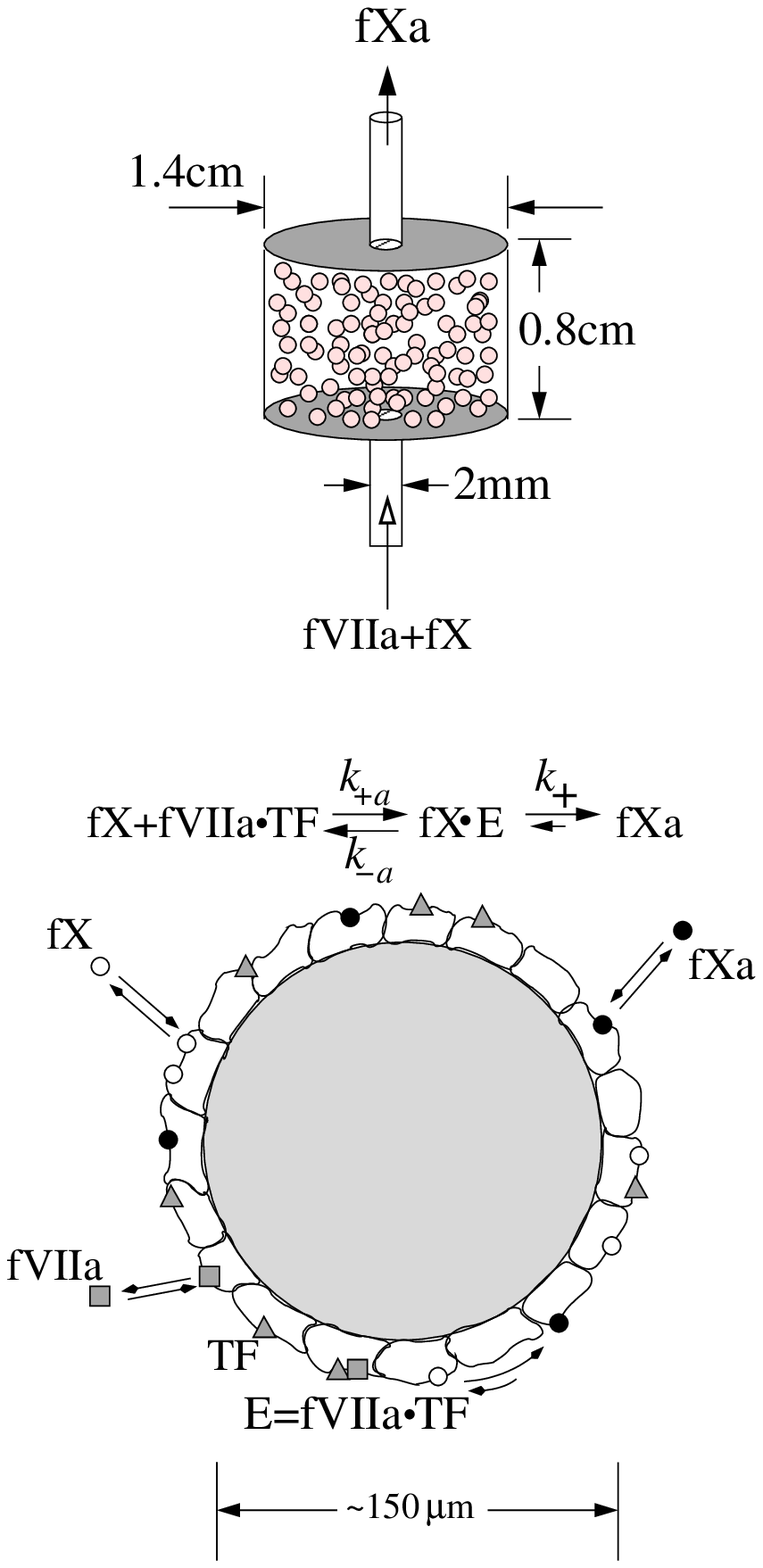}
\end{center}
\vspace{1.4in}
\hspace{5.5in} Fig. 1
\end{figure}

\begin{figure}[htb]
\begin{center}
\leavevmode
\epsfxsize=5.0in
\epsfbox{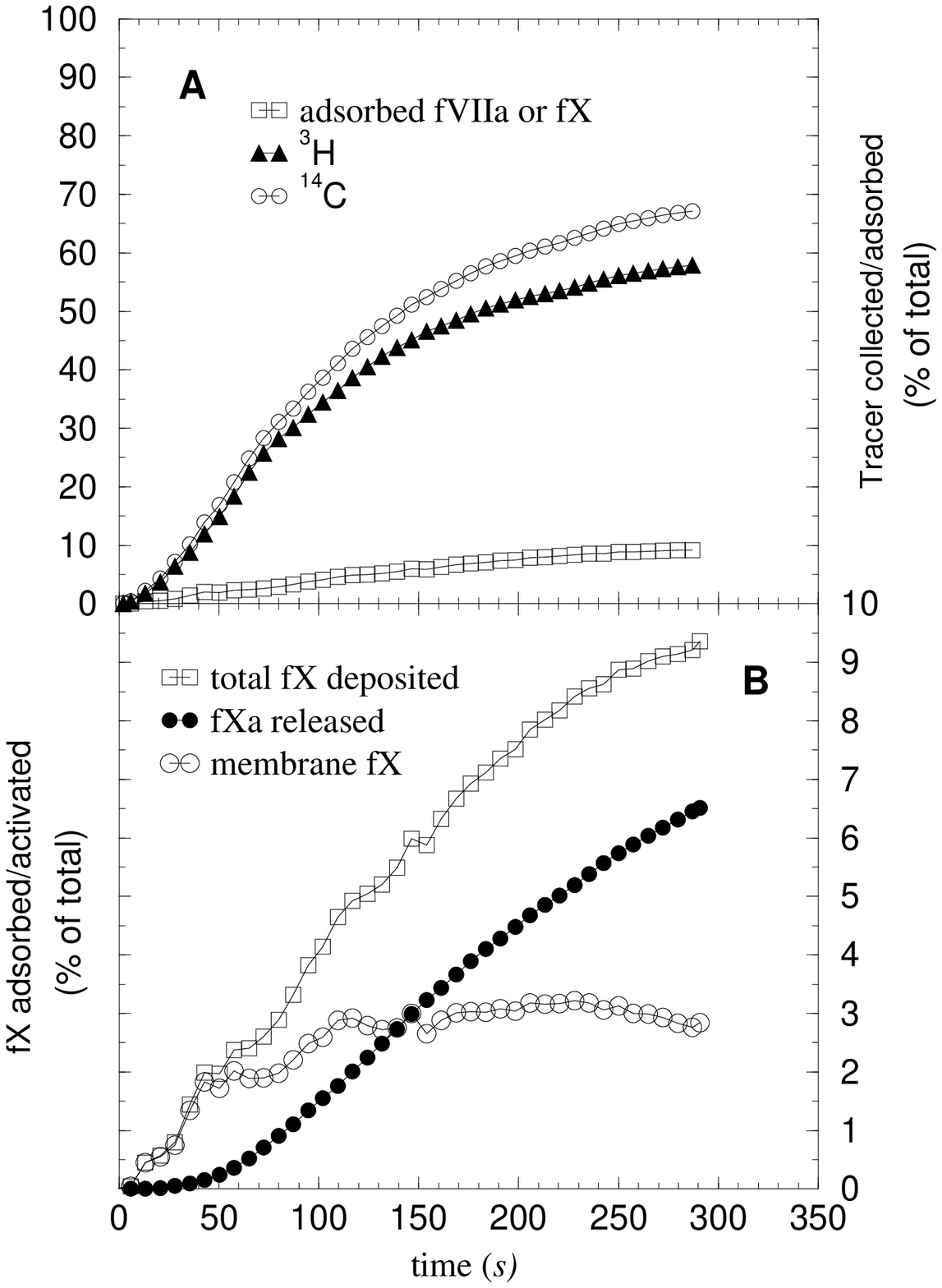}
\end{center}
\vspace{2.5in}
\hspace{5.5in} Fig. 2
\end{figure}

\begin{figure}[htb]
\begin{center}
\leavevmode
\epsfxsize=5.0in
\epsfbox{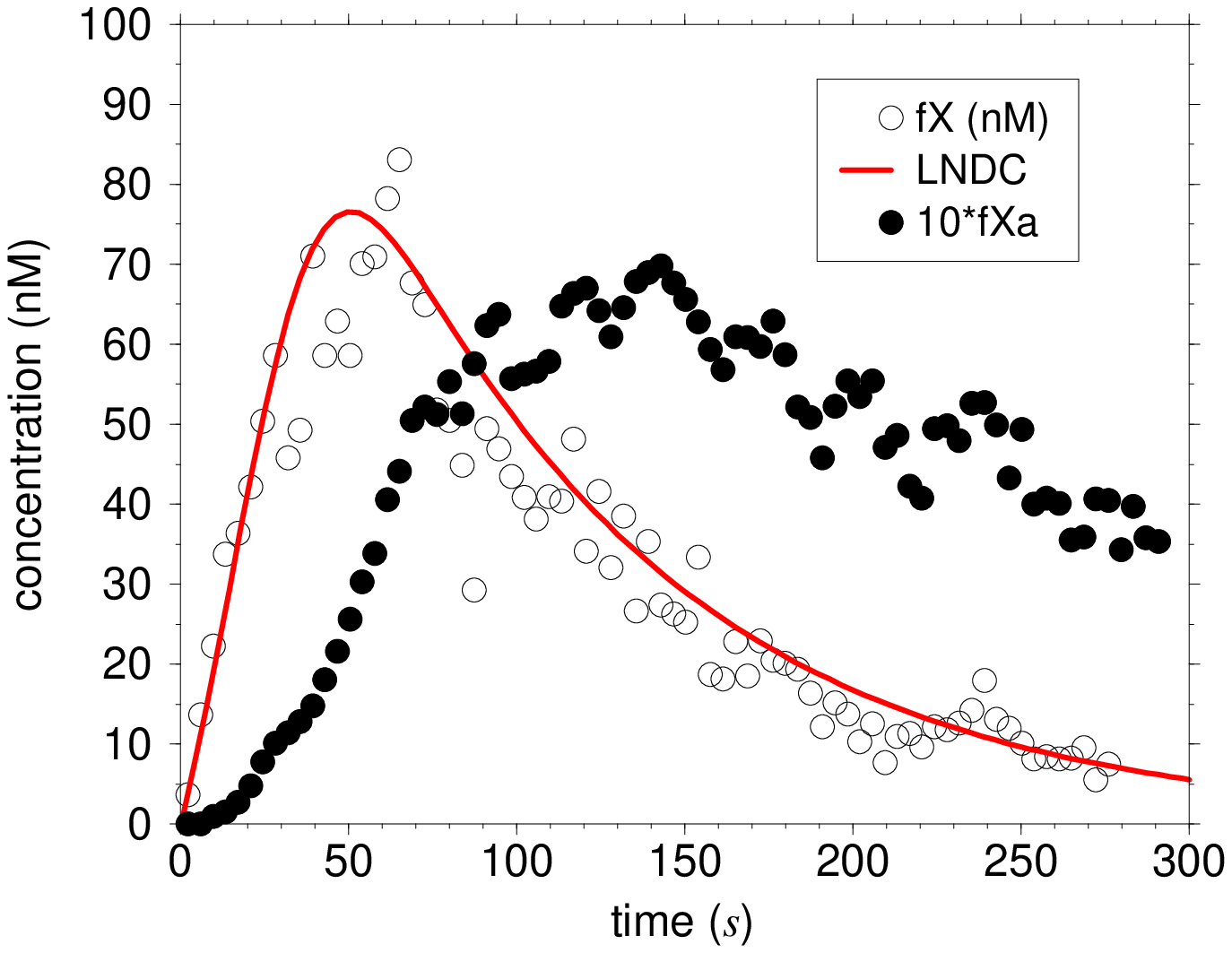}
\end{center}
\vspace{4.2in}
\hspace{5.5in} Fig. 3
\end{figure}

\begin{figure}[htb]
\begin{center}
\leavevmode
\epsfxsize=5.0in
\epsfbox{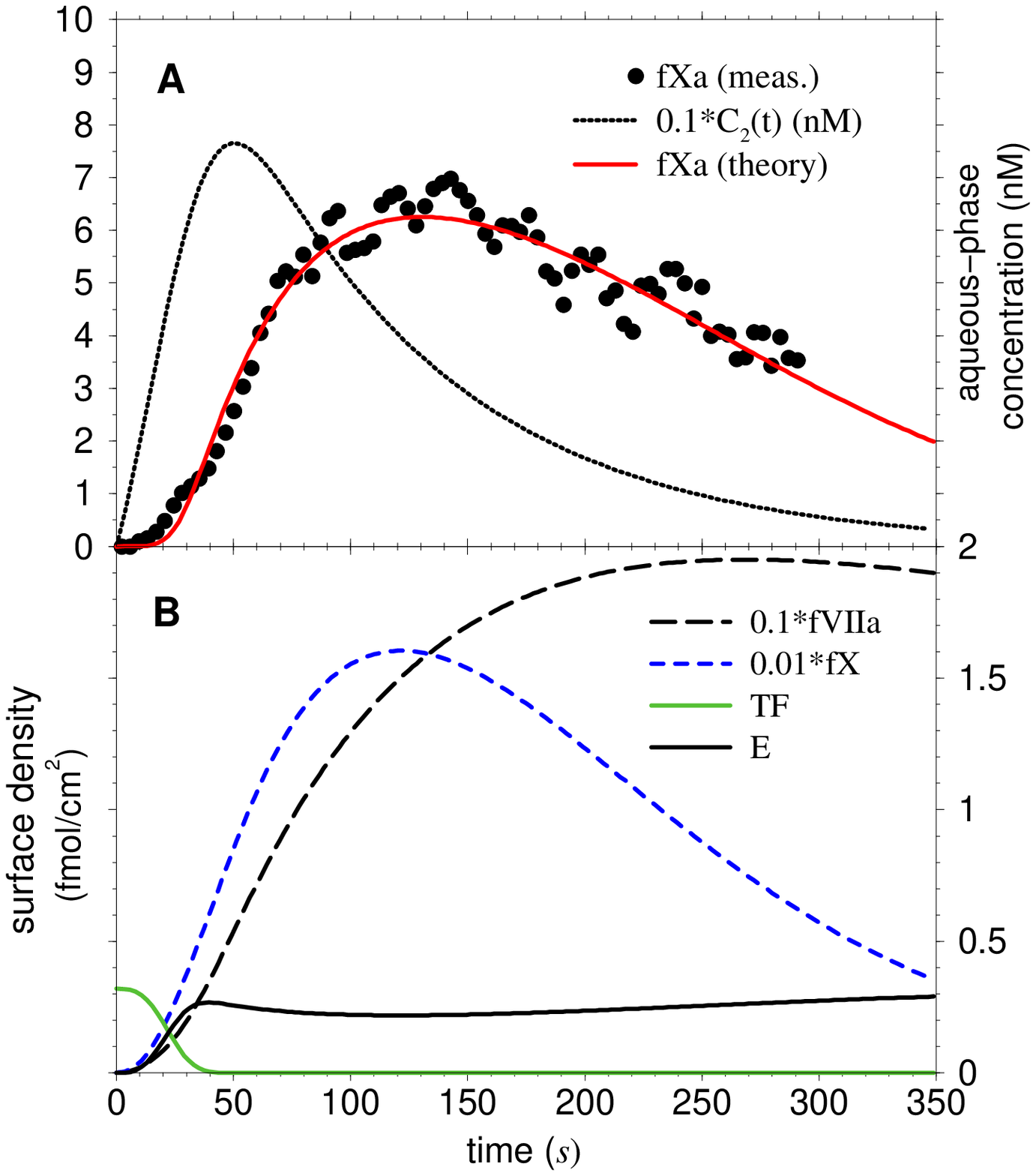}
\end{center}
\vspace{2.8in}
\hspace{5.5in} Fig. 4
\end{figure}

\begin{figure}[htb]
\begin{center}
\leavevmode
\epsfxsize=5.0in
\epsfbox{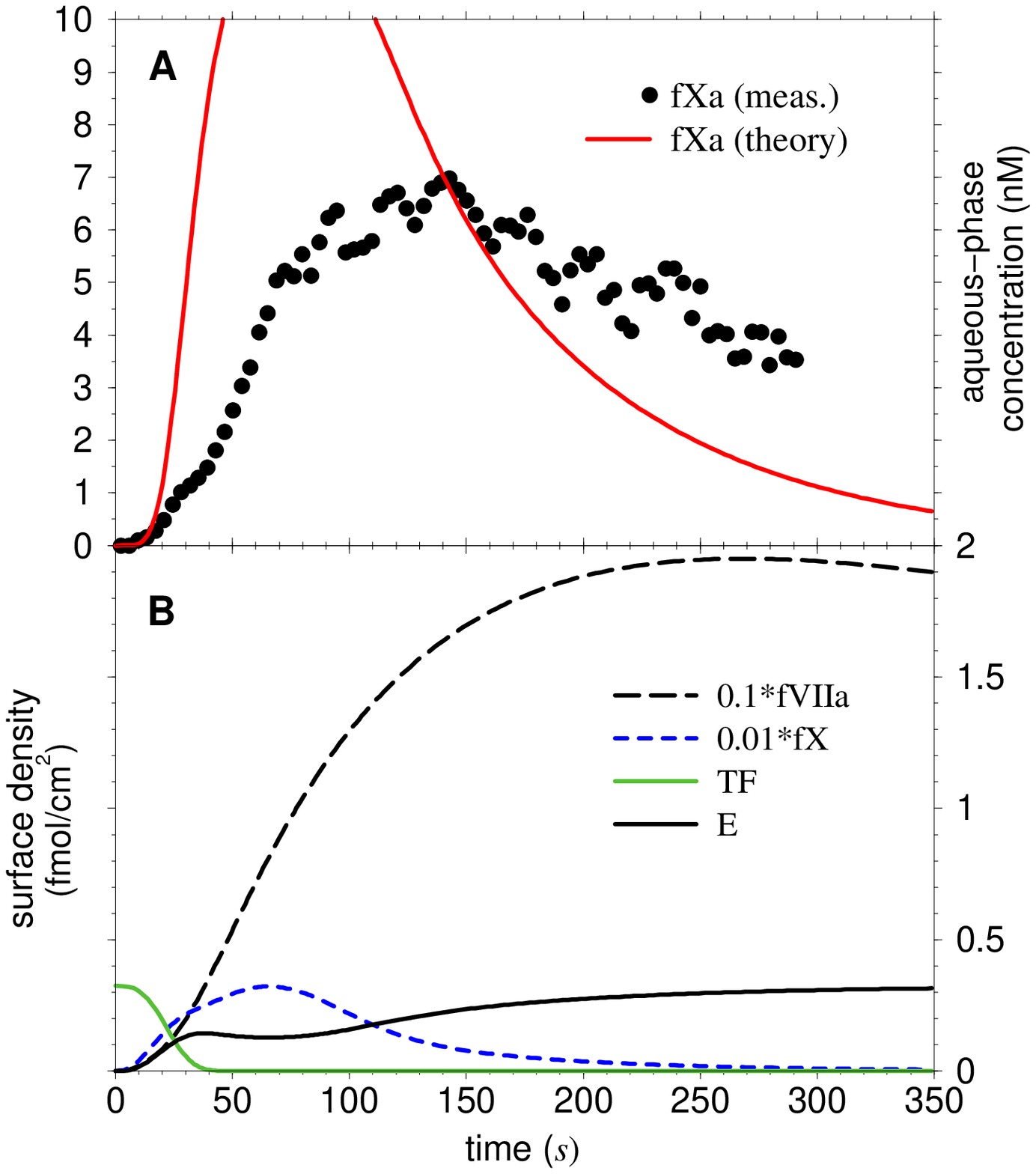}
\end{center}
\vspace{2.8in}
\hspace{5.5in} Fig. 5
\end{figure}

\begin{figure}[htb]
\begin{center}
\leavevmode
\epsfxsize=5.0in
\epsfbox{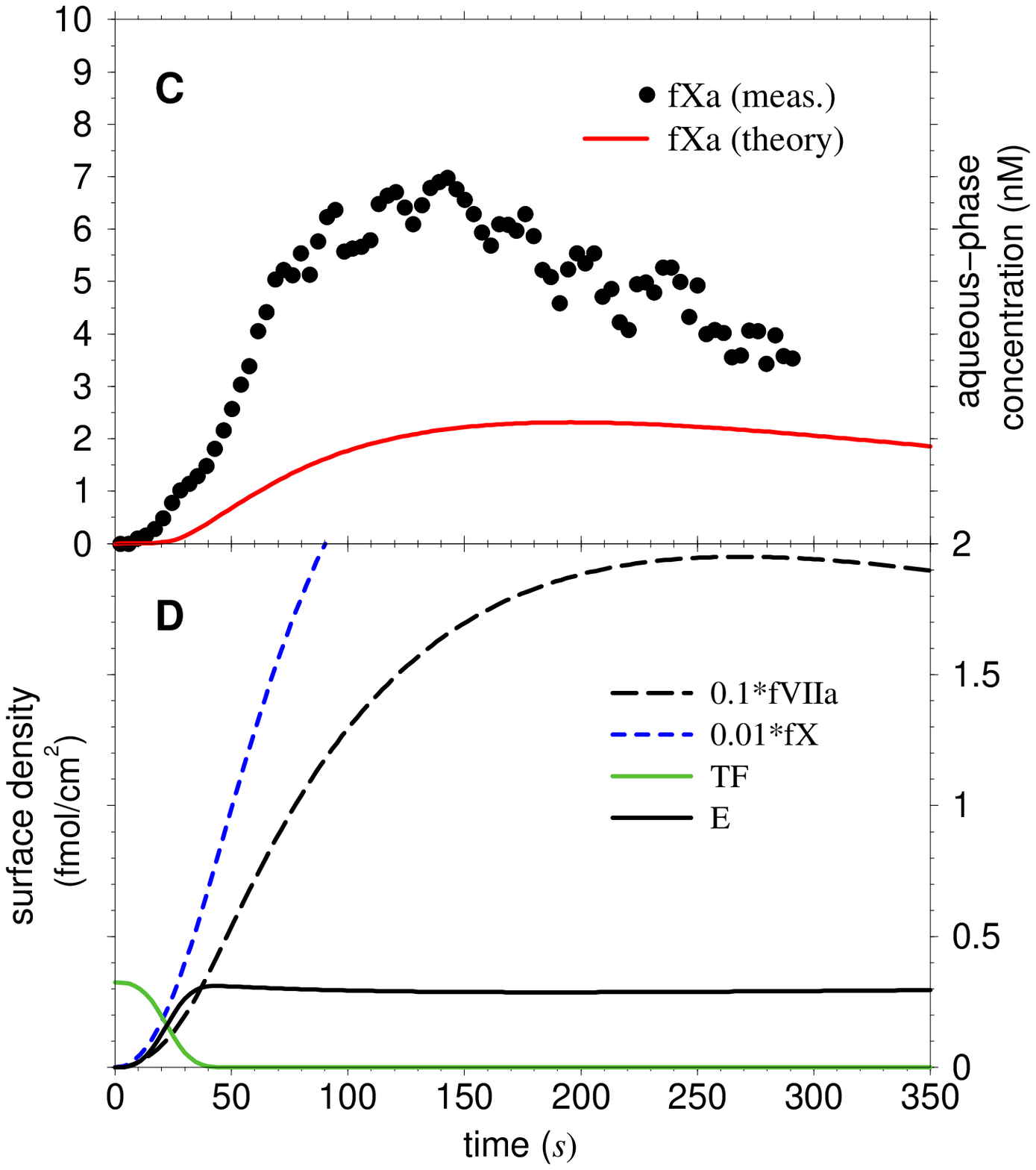}
\end{center}
\vspace{2.8in}
\hspace{5.5in} Fig. 5
\end{figure}



\begin{figure}[htb]
\begin{center}
\leavevmode
\epsfxsize=5.1in
\epsfbox{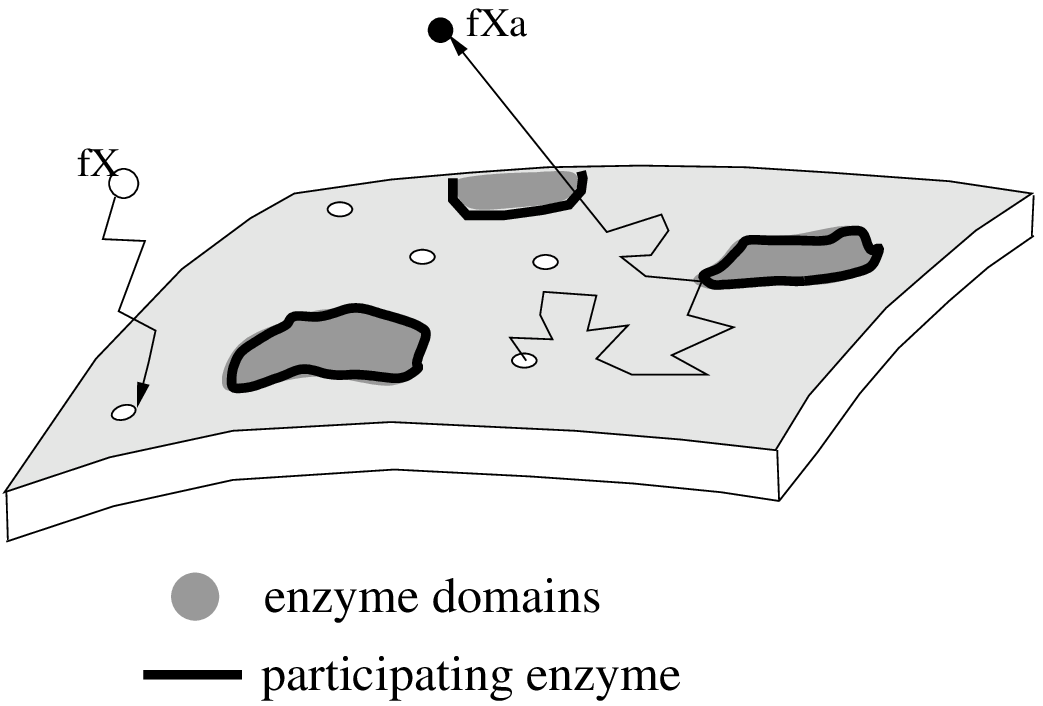}
\end{center}
\vspace{5in}
\hspace{5.5in} Fig. 6
\end{figure}


\begin{thebibliography}{99}

\bibitem[1]{MCF64} 1. Mcfarlane, R.G. (1964) 
{\it Nature} {\bf 202}, 498-499

\bibitem[2]{DOB98} 2. Dobroski, D.R., Rabbani, L.E. and
Loscalzo, Z. (1998) in  {it Thombosis and Hemorrage}
(Loscalzo, J. and Schafer, A., eds), pp. 837-861,
Williams and Wilkins, Baltimore


\bibitem[3]{DVO85} 3. Dvorak, H.N., Senger, D.R., Dvorak,
A.M., Harvey, B.S., and McDonagh, J. (1985)
{\it Science} {\bf 227}, 1059-1061

\bibitem[4]{COL94} 4. Colman, R., Hirsh, J., Marder, V., and
Salzman, E. (eds) (1994) {\it Hemostasis and Thrombosis: Basic
principles and clinical practice} J B. Lippincott Company, 
Philadelphia

\bibitem[5]{FUR88} 5. Furie, B., Furie, B.C. (1988)
{\it Cell} {\bf 53}, 503-518

\bibitem[6]{NEM86} 6. Nemerson, Y., and  Gentry, R.
(1986) {\it Biochemistry} {\bf 25}, 4020-4025 

\bibitem[7]{DZU98} 7. Dzung, T.L., Borgs, P.,
Toneff, T.W., Witte, M.H., and  Rapaport, S.I. 
(1998)  
{\it Am. J. Physiol.} {\bf 274} 
(Heart Cir Physiol 43): H769-H776

\bibitem[8]{CHA95} 8. Chang, P., Aronson, D.L.,
Borenstein, D.G., and Kessler C.M. (1995)
{\it Am. J. Hematol.} {\bf 50}, 79-83

\bibitem[9]{BRI41} 9. Brinkhous, K.M., and Werner, B. (1941)
{\it Am. J. Physiol.} {\bf 132}, 666-669

\bibitem[10]{ZWE79} 10. Zweifach B.W., and Silberberg, A. 
(1979) in {\it Cardiovascular Physiology III}
(Guyton, A.C., and Young, D.B., eds),
pp. 216-260, Baltimore, University Park  

\bibitem[11]{DRA89} 11. Drake, T.A., Morrissey, J.H., and
Edgington, T.S. (1989)
{\it Am. J. Pathol.} {\it 134}, 1087-1097

\bibitem[12]{GRA93} 12. Grabowiski, E., Zuckerman, D.B., and
Nemerson, Y. (1993) {\it Blood} {\bf 81}, 3265-3270

\bibitem[13]{MCG94} 13. McGee, M.P., Foster, S., and Wang, X.
(1994) {\it  J. Exp Med.} {\bf 197}, 1847-1854 

\bibitem[14]{ROT84} 14. Rothberger, H., McGee, M.P., 
and Lee, T.K. (1984) 
{\it J. Clin. Invest.} {\bf 73}, 1524-1531

\bibitem[15]{KRI92} 15. Krishnaswamy, S., Field, K.A., 
Edgington, T.S., Morrissey, J.H., and Mann, K.G. (1992)
{\it J. Biol. Chem.} {\bf 267}, 26110-26120

\bibitem[16]{LON96} 16. London, F.,  and
Walsh, P.N. (1996) 
{\it Biochemistry} {\bf 35}, 12146-12154

\bibitem[17]{MCG98} 17. McGee, M.P., Teuschler,  and H.,  
Liang,  J. (1998) 
{\it Biochem. J.} {\bf 330}, 533-539

\bibitem[18]{MCD97} 18. McDonald, J.F., Shah, A.M.,
Schwalbe, R.A., Kisiel, W.,  Dahlback, B.,  
Nelsestuen, G. (1997) 
{\it Biochemistry} {\bf 36}, 5120-5127

\bibitem[19]{NEL78} 19. Nelsestuen, G.L., Kisiel, W., and  
DiScipio, R.G. (1978)
{\it Biochemistry} {\bf 17}, 2134-2138

\bibitem[20]{RAO88} 20. Rao, L.V.M., and  Rapaport, S.I.
(1988)
{\it Proc. Natl. Acad. Sci. USA} {\bf 85}, 
6687-6691

\bibitem[21]{BAC86} 21. Bach, R., Gentry, R.,  
ad Nemerson, Y. (1986) 
{\it Biochemistry} {\bf 17}, 2134-2138

\bibitem[22]{KOP84} 22. Kop, J.M., Cuypers, P.A.,  
Lindhout, T., Coenraad,  H.,  and Hermens, W.T. (1984)  
{\it J. Biol. Chem.}  {\bf 259}, 1393-1398

\bibitem[23]{SCA96} 23. Scandura, J.M., Ahmad, S.S.,  
and Walsh, P.N. (1996)
{\it Biochemistry} {\bf 35}, 8890-8902

\bibitem[24]{DZU92} 24. Dzung, T.L., Rapaport, S.I.,  
and Rao, V.M. (1992) 
{\it J. Biol. Chem.} {\bf 267}, 15447-15454

\bibitem[25]{PLO67} 25. Ploplis, V.A., Edgington, T.S., 
ad Fair, D.S. (1967) 
{\it J. Biol. Chem.} {\bf 262}, 9503-9508

\bibitem[26]{FAI87} 26. Fair, D.S., MacDonald, M. (1987) 
{\it J. Biol. Chem.} {\bf 262}, 11692-11698

\bibitem[27]{MCG92a} 27. McGee, M.P., Li, L.C., and
Xiong, H. (1992) 
{\it J. Biol. Chem.} {\bf 267}, 24333-24339

\bibitem[28] {ELL98} 28. Ellison, E., and Castellino, F.J. (1998)
{\it Biochemistry} {\bf 37}, 7997-8003

\bibitem[29]{MCG99} 29. McGee, M.P., and Teuchsler, H. (1999)
{\it Thromb. Haemos} {\bf 82}, 93-99 

\bibitem[30]{REP90} 30. Repke, D., Gemmell, C.H., 
Guha, A., Turitto, V.T.,  and Broze, G.J. (1990) 
{\it Proc. Natl. Acad. Sci. USA} {\bf 87}, 
7623-7627

\bibitem[31]{BIL95} 31. Billy, D., Speijer, H.,  Willems,
G., Hemker, H.C., and Lindhou, T. (1995)
{\it J. Biol. Chem.} {\bf 270}, 1029-1034

\bibitem[32]{CRO63} 32. Crone, C. (1963) 
{\it Acta Physiol. Scand.}, {\bf 181}, 103-113

\bibitem[33]{EAT81} 33. Eaton, B.M., and Yudilevich, D.L. 
(1981) 
{\it Am. J. Physiol.} {\bf 241}, C106-112

\bibitem[34]{PER86} 34. Peran, S., and McGee, M.P. (1986)
{\it Biochem. Biophys. Acta} {\bf 856}, 231-236

\bibitem[35]{MCG89} 35. McGee, M.P., Wallin, R., 
Wheeler, F.B., and Rotheberger, H. (1989) {\it Blood},
{\bf 74}, 1583-1590

\bibitem[36]{BAS} 36. Bassingthwaighte, J.B., and Goressky,
C.A. (1984) in {\it A Handbook of Physiology, The
Cardiovascular System, vol. IV: Microcirculation} 
(Renkin, E.M. and  Michel, C.C., eds) pp. 549-626,
Williams and Wilkins, Baltimore

\bibitem[37]{BAS66} 37. 
Bassingthwaighte, J.B., Ackerman, F.H., and
Wood, E.H. (1966) {\it Circulation Res.} {\bf XVIII}, 398.

\bibitem[38]{SCH75} 38. Schmid-Schoenbein, G.W., and 
Zweifach, W. (1975)  
{\it Microvas. Res.} {\bf 10}, 153-164

\bibitem[39]{CAR64} 39. Carson, S.D. (1987)
{\it Thromb. Res.} {\bf 47}, 379-387

\bibitem[40]{VAN71} 40. Van Lenten, L., Ashwell, G. (1971)
{\it J. Biol. Chem.} {\bf 246}, 1889-1894

\bibitem[41]{SAN85} 41. Sanders, N.L., Bajaj, S.P., 
Zivelin, A., and Rapaport, S.T. (1985)
{\it Blood}, {\bf 66}, 204-201

\bibitem[42]{BER83} 42. Berg, C.H. 
Random Walks in Biology. (1983) 
Princeton University Press, Princeton, NJ  

\bibitem[43]{GIR96} 43. Gir, S., Slack, S.M., and Turitto,
V.T. (1996) 
{\it Ann. Biomed. Eng.} {\bf 24}, 394-399

\bibitem[44] {SEA96} 44. Seaton, B.A., and Roberts, M.F. (1996)
in {\it Biological membranes: A molecular 
perspective from computation and experiments}. 
(Merz, K.M., and Roux, B., eds), pp. 355-403,
Birkhauser, Boston, and references therein

\bibitem[45]{CHE79} 45. Cherry, R.J. (1979)
{\it Biochim. Biophys. Acta} {\bf 559}, 289-327

\bibitem[46]{CUR84} 46. Curry, F.E. (1984) in 
{\it Handbook of Physiology, The cardiovascular System, vol IV:
Microcirculation} (Renkin, E.M. and  Michel, C.C., eds) pp.
314-316, Williams and Wilkins, Baltimore

\bibitem[47]{KAR87} 47. Karino, T., Goldsmith, H., 
Motomiya, M., 
Mabuchi, S., and Sohara, Y. (1987) {\it Ann NY 
Acad. Sci.} {\bf 516}, 422-441

\bibitem[48]{AND} 48. Andelman, D.,
Brochard, F., and Joanny, J.-F. (1987)  
{\it J. Chem. Phys.}, {\bf 86}, 3673-3681

\bibitem[49]{MCC} 49. McConnell, H. M. (1991)
{\it Ann. Rev. Phys. Chem.} {\bf 42}, 171-195 

\bibitem[50]{YAN96} 50. Yang, L., and Glasser, M. (1996) 
{\it Biochemistry} {\bf 35}, 16227-13974

\bibitem[51]{LEE98} 51. Lee, A.G. (1998)
{\it Biochim. Biophys.  Acta} {\bf 1376}, 297-318

\bibitem[52]{BAG00} 52. Bagatolli, L.A. and Gratton, E.
(2000) 
{\it Biophys. J.} {\bf 78}, 290-305

\bibitem[53]{MOD98} 53. Nusser, Z., H\'{a}jos, N.,
Somogyi, P. and Mody, I. (1998) {\it Nature}
{\bf 395}, 172-177

\bibitem[54]{FRI} 54. Fritschy, J.M. and Mohler, H.
(1995) {\it J. Comp. Neurology} {\bf 359},
154-194

\bibitem[55]{DISCHER} 55. Discher, D. E. and  
Mohandas, N. (1994)
{\it Biophys. J.} {\bf 71}, 1680-1694

\bibitem[56]{KIM} 56. 
Kim, K.S., Neu, J., and Oster, G. (1998)  
{\it Biophys. J.} {\bf 75}, 2274-2291

\bibitem[57]{PURCELL} 57. Berg, H.C. and Purcell, E.M. (1977)
{\it Biophys. J.} {\bf 20}, 193-219

\end{thebibliography}
\end{document}